# Strong Ultraviolet Pulse From a Newborn Type Ia Supernova

Yi Cao[1], S. R. Kulkarni[1,2], D. Andrew Howell[3,4], Avishay Gal-Yam[5], Mansi M. Kasliwal[6], Stefano Valenti[3,4], J. Johansson[7], R. Amanullah[7], A. Goobar[7], J. Sollerman[8], F. Taddia[8], Assaf Horesh[5], Ilan Sagiv[5], S. Bradley Cenko[9], Peter E. Nugent[10,11], Iair Arcavi[3,12], Jason Surace[13], P. R. Woźniak[14], Daniela I. Moody[14], Umaa D. Rebbapragada[15], Brian D. Bue[15] & Neil Gehrels[9]

[1]*Astronomy Department, California Institute of Technology, Pasadena, CA 91125, USA*

[2]*Caltech Optical Observatories, California Institute of Technology, Pasadena, CA 91125, USA.*

[3]*Las Cumbres Observatory Global Telescope Network, 6740 Cortona Dr., Suite 102, Goleta, CA 93117, USA*

[4]*Department of Physics, University of California, Santa Barbara, CA 93106, USA*

[5]*Department of Particle Physics and Astrophysics, Weizmann Institute of Science, Rehovot 76100, Israel*

[6]*Observatories of the Carnegie Institution for Science, 813 Santa Barbara Street, Pasadena, CA 91101, USA*

[7]*The Oskar Klein Centre, Department of Physics, Stockholm University, SE-106 91 Stockholm, Sweden*

[8]*The Oskar Klein Centre, Department of Astronomy, Stockholm University, SE 106 91 Stockholm, Sweden*

[9]*Astrophysics Science Division, NASA Goddard Space Flight Center, Mail Code 661, Greenbelt, MD 20771, USA*

[10]*Lawrence Berkeley National Laboratory, Berkeley, California 94720, USA*

[11]*Astronomy Department, University of California, Berkeley, 501 Campbell Hall, Berkeley,*

*CA 94720, USA*

[12]*Kavli Institute for Theoretical Physics, University of California, Santa Barbara, CA 93106, USA*

[13]*Spitzer Science Center, California Institute of Technology, Pasadena, CA 91125, USA*

[14]*Los Alamos National Laboratory, Los Alamos, NM 87545, USA*

[15]*Jet Propulsion Laboratory, California Institute of Technology, Pasadena, CA 91109, USA*

**Type Ia supernovae[1] are destructive explosions of carbon oxygen white dwarfs[2,3]. Although they are used empirically to measure cosmological distances[4-6], the nature of their progenitors remains mysterious[3], One of the leading progenitor models, called the single degenerate channel, hypothesizes that a white dwarf accretes matter from a companion star and the resulting increase in its central pressure and temperature ignites thermonuclear explosion[3,7,8]. Here we report observations of strong but declining ultraviolet emission from a Type Ia supernova within four days of its explosion. This emission is consistent with theoretical expectations of collision between material ejected by the supernova and a companion star[9], and therefore provides evidence that some Type Ia supernovae arise from the single degenerate channel.**

On UTC (Coordinated Universal Time) 2014 May 3.29 the intermediate Palomar Transient Factory (iPTF)[10] discovered an optical transient, internally designated as iPTF14atg, in the apparent vicinity of the galaxy IC 831 at a distance of 93.7 megaparsecs (see Methods subsection 'Discovery'). No activity had been detected at the same location in the images taken on the previous night and earlier, indicating that the SN likely exploded between May 2.29 and 3.29. Our follow-up spectroscopic campaign (See Extended Data Table 1 for the observation log) established that iPTF14atg was a Type Ia supernova (SN Ia).

Upon discovery we triggered observations with the Ultraviolet/Optical Telescope (UVOT) and the X-ray Telescope (XRT) onboard the Swift space observatory[11] (observation and data reduction is detailed in Methods subsection 'Data acquisition'; raw measurements are shown in Extended Data Table 2). As can be seen in Figure 1, the UV brightness of iPTF14atg declined substantially in the first two observations. A rough energy flux measure in the UV band is provided by $\nu f_\nu \approx 3\times10^{-13}\,\mathrm{ergs\ cm^{-2}\ s^{-1}}$ in the *uvm2* band. Starting from the third epoch, the UV and optical emission began to rise again in a manner similar to that seen in other SNe Ia. The XRT did not detect any X-ray signal at any epoch (Methods subsection 'Data acquisition'). We thus conclude that iPTF 14atg emitted a pulse of radiation primarily in the UV band. This pulse with an observed luminosity of $L_{UV} \approx 3\times10^{41}\,\mathrm{ergs\ s^{-1}}$ was probably already declining by the first epoch of the Swift observations (within four days of its explosion).

Figure 1 also illustrates that such an early UV pulse from a SN Ia within four days of its explosion is unprecedented[12,13]. We now seek an explanation for this early UV emission. As detailed in Methods subsection 'Spherical models for the early UV pulse', we explored models in which the UV emission is spherically symmetric with the SN explosion (such as shock cooling and circumstellar interaction). These models are unable to explain the observed UV pulse. Therefore we turn to asymmetric models in which the UV emission comes from particular directions.

A reasonable physical model is UV emission arising in the ejecta as the ejecta encounters a companion[9, 14]. When the rapidly moving ejecta slams into the companion, a strong reverse shock is generated in the ejecta that heats up the surrounding material. Thermal radiation from the hot material, which peaks in the ultraviolet, can then be seen for a few days until the fast-moving ejecta engulfs the companion and hides the reverse shock region. We compare a semi-analytical model[9] to the *Swift*/UVOT lightcurves. For simplicity, we fix the explosion date at May 3. We assume that the exploding white dwarf is close to the Chandrasekhar mass limit (1.4 solar mass) and that the SN explosion energy is $10^{51}$ ergs. These values lead to a mean expansion velocity of $10^4$ km $s^{-1}$ for the ejecta. Since the temperature at the collision location is so high that most atoms are ionized, the opacity is probably dominated by electron scattering. To further simplify the case, we assume that the emission from the reverse shock region is blackbody and isotropic. In order to explain the UV lightcurves, the companion star should be located 60 solar radii away from the white dwarf (Model A; black dashed curves in Figure 1).

There are several caveats in this simple semi-analytical model. First, the model parameters are degenerate. For example, if we reduce the SN energy by a factor of two and increase the binary separation to 90 solar radii, the model lightcurve can still account for the observed UV luminosities (Model B; blue dashed curves in Figure 1). Second, the emission from the reverse shock region is not isotropic. The UV photons can only easily escape through the conical hole carved out by the companion star and therefore the emission is more concentrated in this direction. Third, the actual explosion date is not well constrained, so that when the companion collision happened is not clear. Our multi-wavelength observations soon after discovery of the SN provide a good data set for detailed modeling.

We also construct the spectral energy distribution (SED) from the photometry and spectrum of iPTF14atg obtained on the same day of the first UVOT epoch and compare it with the blackbody spectra derived from Models A and B. As can be seen in Figure 2, the model blackbody spectra are consistent with the overall shape of the SED, indicating that the emitting regions can be approximated by a blackbody with a temperature of 11000 K and a radius of 6,000 solar radii.

Next, given the diversity of Type Ia SNe, we investigate the specifics of iPTF14atg using its multi-band lightcurves (Figure 3) and spectra (Figure 4). First, the existence of Si II and S II absorption features in the pre-maximum spectra indicates that iPTF14atg is spectroscopically a Type Ia SN[1]. Second, iPTF14atg, with a peak absolute magnitude of −17.9 mag in the *B*-band, is 1.4 magnitudes fainter than normal SNe Ia that are used as cosmological distance indicators[15]. Subluminous SNe Ia belong to three major families with prototypical events being SN1991bg[16], SN2002cx[17, 18], and SN2002es[19]. A comparative analysis of lightcurves and spectra between iPTF14atg and the three families (detailed in Methods subsection 'SN specification') shows that: iPTF14atg is more luminous than SN1991bg and evolves more slowly than SN1991bg in both the rise and decline phases. The expansion velocity of iPTF14atg estimated from absorption lines are systematically lower than those of SN1991bg. SN2002cx and iPTF14atg have similar lightcurves, but iPTF14atg shows deep absorption features in the pre-maximum spectra that are not seen in SN2002cx and the post-maximum absorption features of iPTF14atg are generally weaker than those seen in SN2002cx. We only have limited knowledge about the evolution of SN2002es. Despite the fact that SN2002es is one magnitude brighter at peak than iPTF14atg and that the lightcurve of SN2002es shows an accelerating decline about 30 days after its peak, which is not seen in iPTF14atg, iPTF14atg shows a reasonable match to SN2002es in both lightcurve shape and spectra with higher line velocities. In addition, the host galaxy IC831 of iPTF14atg is an early-type galaxy. This is consistent with hosts of known SN2002es-like events, while the majority of SN2002cx-like events occur in late-type galaxies[18, 20]. Therefore, we tentatively classify iPTF14atg as a high-velocity version of SN2002es.

Our work along with recent suggestions for companions in SN2008ha[21] and SN2012Z[22] hints that subluminous SNe with low velocities, such as SN2002cx and SN2002es, arise from the single degenerate channel. In contrast, there is mounting evidence that some SNe Ia result from the double degenerate channel[23, 24] in which two WDs merge or collide and then explode in a binary or even triple system[3]. Clearly determining the fraction of SNe Ia with companion interaction signatures can disentangle the SN Ia progenitor puzzle. Prior to our discovery, searches for companion interaction have been carried out in both UV[13, 25, 26] and ground-based optical data[27, 28]. However, very few SNe were observed in the UV within a few days of their explosions. Our observation of iPTF14atg also demonstrates that the interaction signature is not distinctive in the optical bands.

Therefore rapid UV follow-up observations of extremely young SNe or fast-cadence UV transient surveys are warranted to probe the companion interaction of SNe Ia. Given the observed UV flux of iPTF14atg, ULTRASAT[29] (a proposed space telescope aimed at undertaking fast-cadence observations of the UV sky) should detect such events up to 300 megaparsecs. Factoring in its field-of-view of 210 square degrees, ULTRASAT will detect three dozen SNe Ia of all kinds within this volume during its two-year mission lifetime. In fact, the UV flux of the SN-companion interaction is brighter at earlier phases. Thus, ULTRASAT may discover more such events at greater distances. Since up to a third of SNe Ia are subluminous[18], ULTRASAT survey could definitively determine the fraction of events with companion interaction and thus the rate of events from the single degenerate channel.

1. Filippenko, A. V. Optical Spectra of Supernovae. *Ann. Rev. Astron. Astrophys.* **35**, 309–355 (1997)

2. Nugent, P. E. *et al.* Supernova SN 2011fe from an exploding carbon-oxygen white dwarf star. *Nature* **480**, 344–347 (2011).

3. Maoz, D., Mannucci, F. & Nelemans, G. Observational Clues to the Progenitors of Type Ia Supernovae. *Ann. Rev. Astron. Astrophys.* **52**, 107–170 (2014).

4. Riess, A. G. *et al.* Observational Evidence from Supernovae for an Accelerating Universe and a Cosmological Constant. *Astron. J.* **116**, 1009–1038 (1998).

5. Perlmutter, S. *et al.* Measurements of Ω and Λ from 42 High-Redshift Supernovae. *Astrophys. J.* **517**, 565–586 (1999).

6. Sullivan, M. *et al.* SNLS3: Constraints on Dark Energy Combining the Supernova Legacy Survey Three-year Data with Other Probes. *Astrophys. J.* **737**, 102 (2011).

7. Whelan, J. & Iben, I., Jr. Binaries and Supernovae of Type I. *Astrophys. J.* **186**, 1007–1014 (1973).

8. Wang, B. & Han, Z. Progenitors of type Ia supernovae. *New Astronomy Reviews* **56**, 122–141 (2012).

9. Kasen, D. Seeing the Collision of a Supernova with Its Companion Star. *Astrophys. J.* **708**, 1025–1031 (2010).

10. Law, N. M. *et al.* The Palomar Transient Factory: System Overview, Performance, and First Results. *Publ. Astron. Soc. Pac.* **121**, 1395–1408 (2009).

11. Gehrels, N. *et al.* The Swift Gamma-Ray Burst Mission. *Astrophys. J.* **611**, 1005–1020 (2004).

12. Milne, P. A. *et al.* Near-ultraviolet Properties of a Large Sample of Type Ia Supernovae as Observed with the Swift UVOT. *Astrophys. J.* **721**, 1627–1655 (2010).

13. Brown, P. J. *et al.* A Swift Look at SN 2011fe: The Earliest Ultraviolet Observations of a Type Ia Supernova. *Astrophys. J.* **753**, 22 (2012).

14. Pan, K.-C., Ricker, P. M. & Taam, R. E. Impact of Type Ia Supernova Ejecta on Binary Companions in the Single-degenerate Scenario. *Astrophys. J.* **750**, 151 (2012).

15. Yasuda, N. & Fukugita, M. Luminosity Functions of Type Ia Supernovae and Their Host Galaxies from the Sloan Digital Sky Survey. *Astron. J.* **139**, 39–52 (2010).

16. Filippenko, A. V. *et al.* The subluminous, spectroscopically peculiar type IA supernova 1991bg in the elliptical galaxy NGC 4374. *Astron. J.* **104**, 1543–1556 (1992).

17. Li, W. *et al.* SN 2002cx: The Most Peculiar Known Type Ia Supernova. *Publ. Astron. Soc. Pac.* **115**, 453–473 (2003).

18. Foley, R. J. *et al.* Type Iax Supernovae: A New Class of Stellar Explosion. *Astrophys. J.* **767**, 57 (2013).

19. Ganeshalingam, M. *et al.* The Low-velocity, Rapidly Fading Type Ia Supernova 2002es. *Astrophys. J.* **751**, 142 (2012).

20. White, C. J. *et al.* Slow-speed Supernovae from the Palomar Transient Factory: Two Channels. *Astrophys. J.* **799**, 52 (2015).

21. Foley, R. J. *et al.* Possible Detection of the Stellar Donor or Remnant for the Type Iax Supernova 2008ha. *Astrophys. J.* **792**, 29 (2014).

22. McCully, C. *et al.* A luminous, blue progenitor system for the type Iax supernova 2012Z. *Nature* **512**, 54–56 (2014).

23. Li, W. *et al.* Exclusion of a luminous red giant as a companion star to the progenitor of supernova SN 2011fe. *Nature* **480**, 348–350 (2011).

24. González Hernández, J. I. *et al.* No surviving evolved companions of the progenitor of SN 1006. *Nature* **489**, 533–536 (2012).

25. Foley, R. J. *et al.* Very Early Ultraviolet and Optical Observations of the Type Ia Supernova 2009ig. *Astrophys. J.* **744**, 38 (2012).

26. Brown, P. J. *et al.* Constraints on Type Ia Supernova Progenitor Companions from Early Ultraviolet Observations with Swift. *Astrophys. J.* **749**, 18 (2012).

27. Hayden, B. T. *et al.* Single or Double Degenerate Progenitors? Searching for Shock Emission in the SDSS-II Type Ia Supernovae. *Astrophys. J.* **722**, 1691–1698 (2010).

28. Bianco, F. B. *et al.* Constraining Type Ia Supernovae Progenitors from Three Years of Supernova Legacy Survey Data. *Astrophys. J.* **741**, 20 (2011).

29. Sagiv, I. *et al.* Science with a Wide-field UV Transient Explorer. *Astron. J.* **147**, 79 (2014).

**Acknowledgements** We thank A. L. Piro, M. Kromer and J. Cohen for helpful discussion. We also thank A. Waszczak, A. Rubin, O. Yaron, A. De Cia, D. Perley, G. Duggan, O. Smirnova, S. Papadogiannakis, A. Nyholm, Y. F. Martinez and the staff at the Nordic Optical Telescope and Gemini for observation and data reduction. Some of the data presented here were obtained at the W. M. Keck Observatory, which is operated as a scientific partnership among the California Institute of Technology, the University of California and NASA. The observatory was made possible by the generous financial support of the W. M. Keck Foundation. Some data were obtained with the Nordic Optical Telescope, which is operated by the Nordic Optical Telescope Scientific Association at the Observatorio del Roque de los Muchachos, La Palma, Spain. This work also makes use of observations from the LCOGT network. Research at California Institute of Technology is supported by the National Science Foundation. D.A.H. acknowledges support from the National Science Foundation. A.G.-Y. acknowledges support from the EU/FP7 via an ERC grant, the "Quantum Universe" I-Core program, ISF, Minerva and Weizmann-UK grants, and the Kimmel Award. M.M.K. acknowledges generous support from the Carnegie-Princeton fellowship. Supernova research at the OKC is supported by the Swedish Research Council and by the Knut and Alice Wallenberg Foundation. The National Energy Research Scientific Computing Center, which is supported by the Office of Science of the US Department of Energy under contract No. DE-AC02-05CH11231, provided staff, computational resources, and data storage for this project. LANL participation in iPTF is supported by the US Department of Energy as part of the Laboratory Directed Research and Development program. A portion of this work was carried out at the Jet Propulsion Laboratory under a Research and Technology

Development Grant, under contract with the National Aeronautics and Space Administration.

**Author Contributions** Y.C. initiated the study, conducted analysis and wrote the manuscript. S.R.K. is iPTF PI, and contributed to *Swift*/UVOT data analysis and manuscript preparation. S.V. and D.A.H. contributed to LCOGT observation, data analysis and manuscript preparation. A.G.-Y. contributed to manuscript preparation. M.M.K. contributed to *Swift*, APO and Gemini-N observations and manuscript preparation. J.J., A.G., J.S., F.T. and R.A. triggered early NOT observation and contributed to manuscript preparation. A.H. triggered JVLA observation and analyzed the data. I.S. found the supernova. S.B.C. reduced the P48 data and contributed to manuscript preparation. P.E.N. contributed to manuscript preparation. J.S. and I.A. contributed to building transient vetting and data archiving software. B.D.B., D.I.M., U.D.R. and P.R.W. contributed to the machine learning codes to search for young transients. N.G. is *Swift* PI.

**Author Information** Reprints and permissions information is available at www.nature.com/reprints. The authors declare competing financial interests. Correspondence and requests for materials should be addressed to Y.C. (email: ycao@astro.caltech.edu).

**Figure 1: *Swift*/UVOT lightcurves of iPTF14atg.** iPTF14atg lightcurvs are shown in red circles and lines and are compared with those of other SNe Ia (gray circles). The magnitudes are in the AB system. The 1-σ error bars include both statistical and systematic uncertainties in measurements. Lightcurves of other SNe and their explosion dates are taken from previous studies[13,26]. In each of the three UV bands (*uvw2*, *uvm2* and *uvw1*), iPTF14atg stands out for exhibiting a decaying flux at early times. The blue and black dashed curves show two theoretical lightcurves derived from companion interaction models[9].

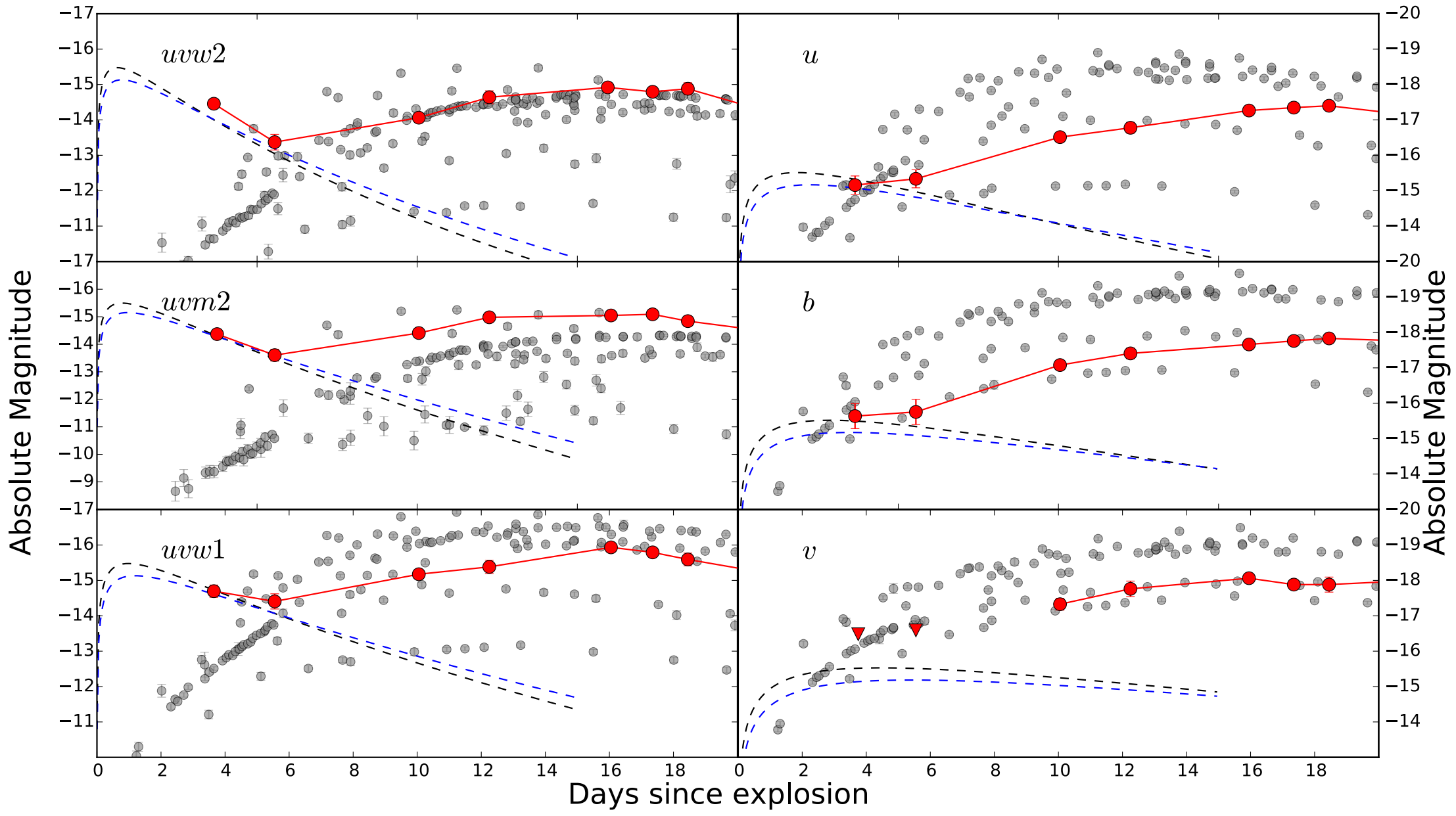

**Figure 2: The spectral energy distribution of iPTF14atg.** The spectral energy distribution of iPTF14atg on May 6 (three days after explosion) is constructed by using the PTF *r*-band magnitude (red), an optical spectrum (gray), and *Swift*/UVOT measurements (green circle) and upper limit (green triangle). The error bars denote 1-σ uncertainties. The blue and black blackbody spectra correspond to the model lightcurves of the same color in Figure 1.

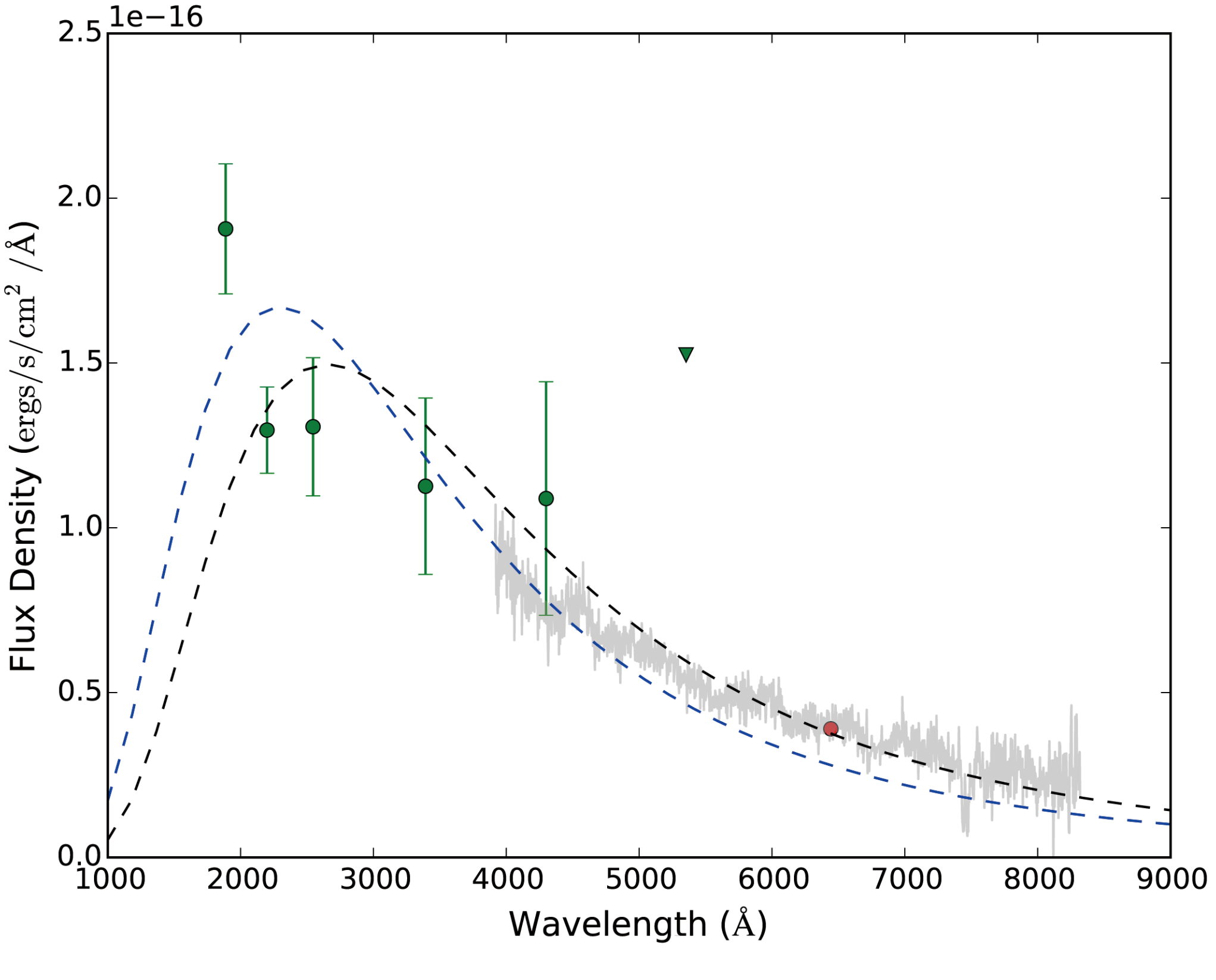

**Figure 3: The multi-color lightcurve of iPTF14atg.** Following the convention, the magnitudes in B and V bands are in the Vega system while those in *g*, *r* and *i* in the AB system. Error bars represent 1-σ uncertainties.

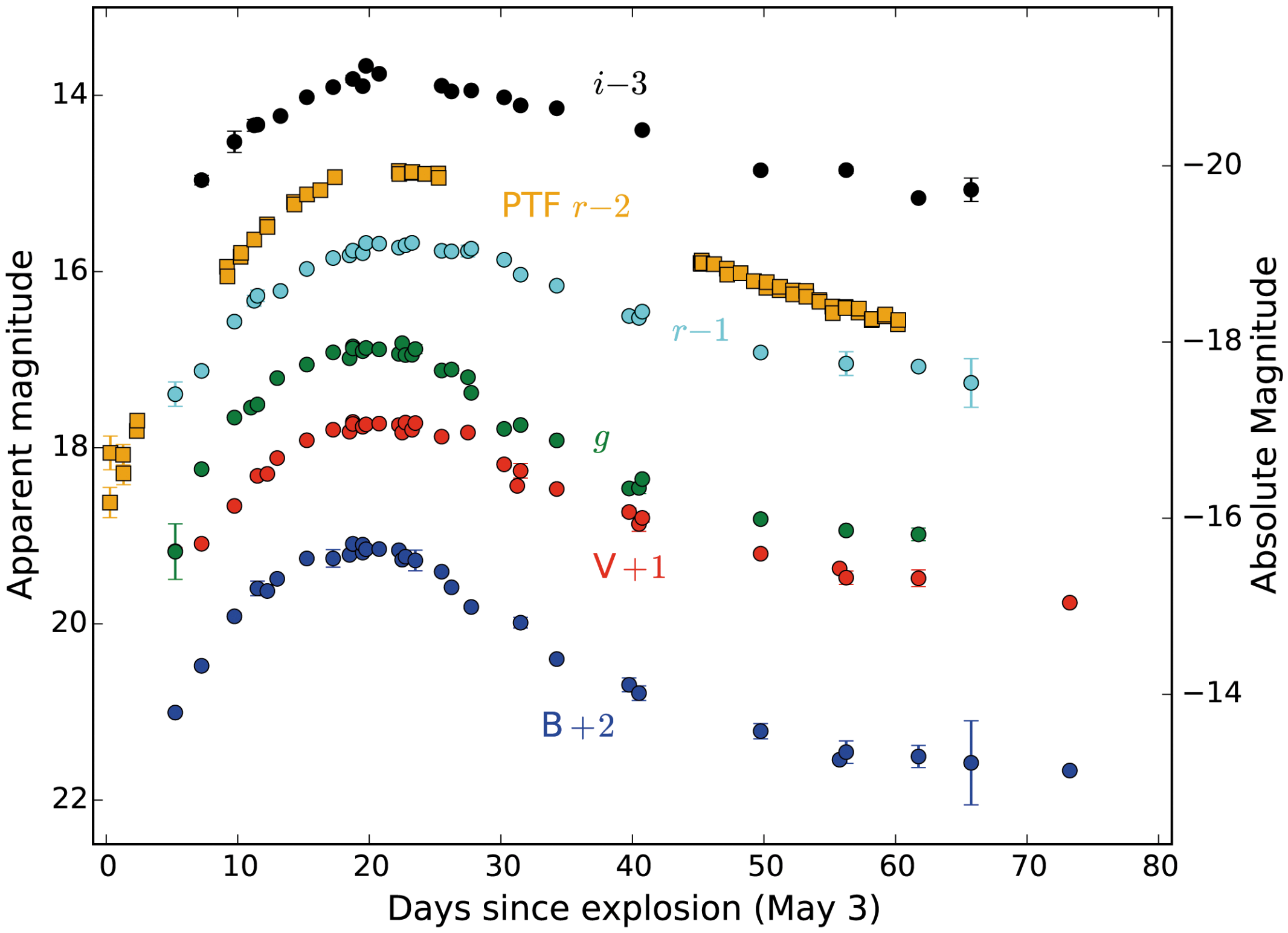

**Figure 4: Spectral evolution of iPTF14atg.** The spectra of iPTF14atg (black) are compared with those of SN2002es (green) at the SN max and a week after max. The SN2002es spectrum at max is blue-shifted by 2,000 km $s^{-1}$ and that at +1 week is blue-shifted by 1,000 km $s^{-1}$. Ticks at the bottom of the plot label major absorption features in the spectra.

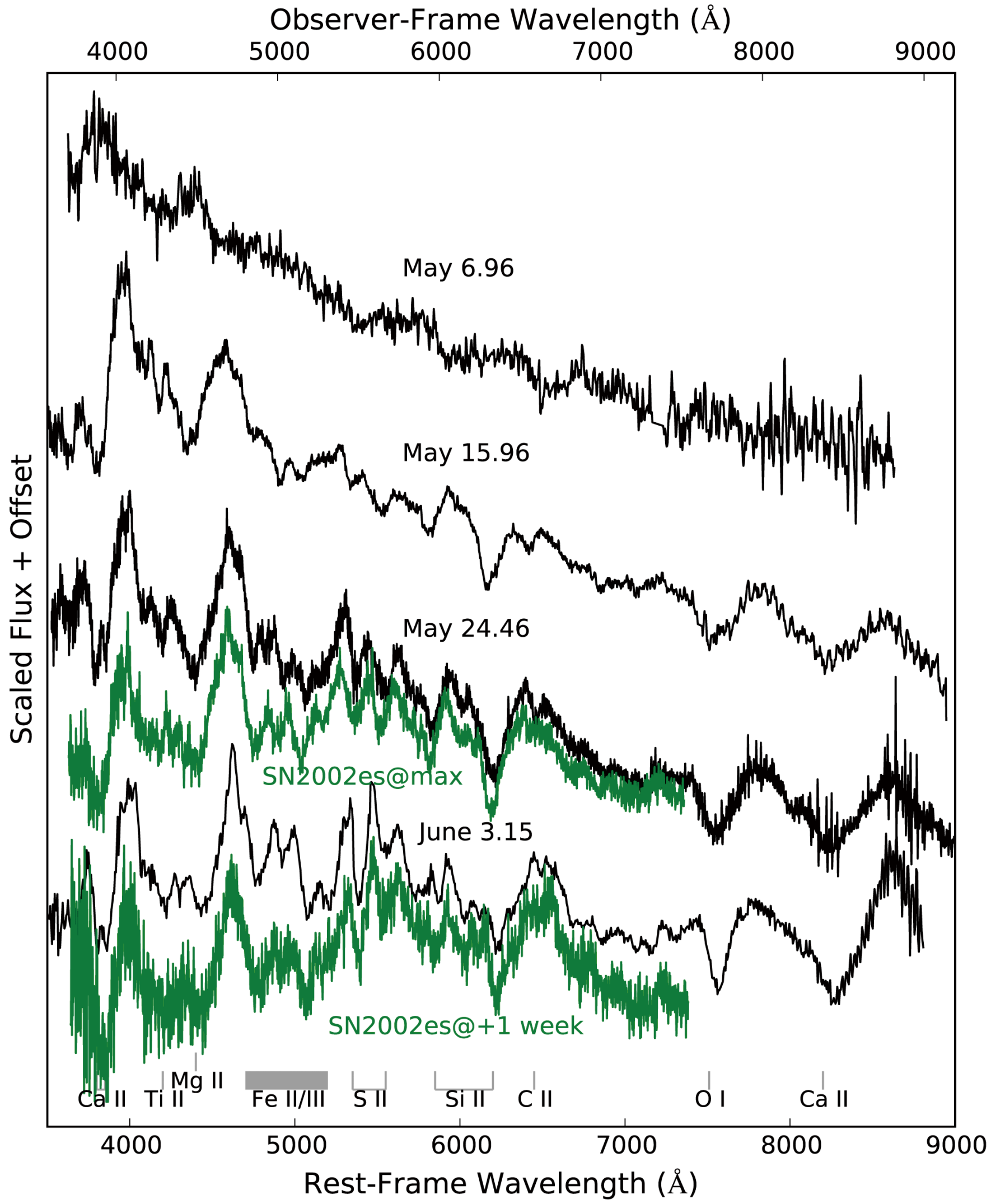

## Methods

### 1 Discovery

The intermediate Palomar Transient Factory (iPTF) uses the 48-inch Samuel Oschin telescope (P48) at Palomar Observatory, CA, USA to characterize optical transients and variable stars[10]. A single P48 frame has a field of view of 7.2 square degrees and achieves a detection threshold of $r(\mathrm{AB}) \sim 21\,\mathrm{mag}$ (5-$\sigma$, i.e., 99.9999% Confidence Level, hereafter 99.9999% CL). In 2014 spring and summer, iPTF conducted a fast-cadence experiment to search for young transients. Every field was monitored twice separated by about an hour every night (weather permitting). Transients were identified in real time by a monitoring group aided by machine-learning classifiers[34–36]. Panchromatic follow-up of young transients was carried out within hours after discovery[37].

The SN iPTF14atg was discovered on UT 2014 May 3.29 at $\alpha$(J2000) = $12^h52^m44^s.8$ and $\delta$(J2000) = +26°28'13", about 10" east with no measurable offset in declination from the apparent host galaxy IC 831. It had an $r$-band magnitude of 20.3 upon discovery. No source was detected at the same location on images taken on UT 2014 May 02.25 and 02.29 down to a limiting magnitude of $r \approx 21.4$ (99.9999% CL). No activity had been found at this location in the iPTF archival data in 2013 (3 epochs) and 2014 (101 epochs) down to similar limiting magnitudes. This SN was also independently discovered by ASASSN on May 22[38] and classified as a SN 1991bg-like Type Ia SN on June 3[39].

SDSS (Data Release 12) measured the redshift of IC 831 to be 0.02129[40]. With the cosmological parameters measured by Planck ($H_0$ = 67.8 km s$^{-1}$ Mpc$^{-1}$, $\Omega_m$ = 0.307, $\Omega_\Lambda$ =

0.691 and $\Omega_\nu$ = 0.001)[41], we calculate a co-moving distance of 93.7 megaparsecs and a distance modulus of 34.9 mag for IC831.

Upon publication, photometric data presented in this paper will be made publicly available through the online version of the letter. Both photometric and spectroscopic data will also be made publicly available on WISeREP[42].

**2 Data acquisition**

***Swift* observations and data reduction** Starting on May 6, the Ultraviolet & Optical Telescope (UVOT)[43] and the X-ray Telescope (XRT)[44] onboard the *Swift* space Observatory[11] observed iPTF14atg for fourteen epochs in May and June (summarized in Extended Data Table 2). In order to subtract the host galaxy contamination, reference images were taken six months after the SN explosion. Visual inspection to the reference images ensures that the SN has faded away.

Photometric measurements of the UVOT images were undertaken with the *uvotsource* routine in the HEASoft package[45]. Instrumental fluxes of iPTF14atg were extracted with an aperture of radius 3” centered at the location determined by the iPTF optical images and the sky background is calculated with an aperture of radius 20” in the vicinity of iPTF14atg. The fluxes were then corrected by the growth curves of UVOT point spread functions (PSFs) and for coincidence loss. Then the instrumental fluxes were converted to physical fluxes using the most recent calibration[46]. The host galaxy flux is measured with the same aperture in the reference images. The XRT data were analyzed with the *Ximage* software in the HEASoft package. We estimated count rate upper limits at a 99.7% CL at the location of iPTF14atg for non-detections.

We use WebPIMMS[47] to convert the XRT upper limit on May 6 to physical quantities. As shown in Figure 2, the optical and UV data taken on the same day can be approximated by a blackbody model with a temperature $T = 1.1 \times 10^4\,\mathrm{K}$ and radius of 6,000 solar radii. Using the blackbody model with the radius fixed to the above value and setting the interstellar column density of hydrogen to $N_H = 8 \times 10^{19}\,\mathrm{cm}^{-2}$ as is appropriate towards this direction[48] we find that the XRT upper limit of counting rate agrees with a temperature $T < 10^5\,\mathrm{K}$.

**Ground-based observations and data reduction** As described in §1, P48 observed the iPTF 14atg field every night (weather-permitting) until July 2. The host galaxy contamination in these images was removed with the aid of a reference image which was built by stacking twelve P48 pre-SN images. We then performed the PSF photometry on the subtracted images. The photometry is calibrated to the PTF-IPAC catalog[49].

We also triggered LCOGT to follow up iPTF14atg in *griBV* filters. Because reference images were not available, we used an image-based model composed of a PSF and a low-order polynomial to model the SN light and its underlying galaxy background simultaneously. The photometry is then calibrated to the SDSS catalog.

Our optical spectroscopic observation log is presented in Extended Data Table 1. Spectroscopic data were reduced with standard routines in IRAF and/or IDL.

We also observed iPTF14atg with the Jansky Very Large Array on May 16 at both 6.1 GHz (C-band) and 22 GHz (K-band). The observation was performed in the A configuration using J1310+3220 as a phase calibrator and 3C286 as a flux calibrator. The data were reduced using standard routines in the CASA software. The observation resulted in a null-detection at both bands with an upper limit of 30 $\mu$Jy (99.7% CL). Note that the radio observation is taken ten days after iPTF14atg discovery.

**3 Spherical models for the early UV pulse**

In this section we consider models spherically symmetric to the SN explosion center to interpret the early UV excess seen in iPTF14atg. In the first model, we investigate the possibility that the UV pulse is powered by radioactive decay. The rise time of a SN peak is roughly characterized by the diffusive timescale from the radioactive layer to the photosphere, i.e.,

$$\tau \propto \left( \frac{\kappa M_{ej}}{c v_{\exp}} \right)^{1/2},$$

where $M_{ej}$ is the mass of ejecta outside the radioactive layer, $\kappa$ is the mean opacity of the ejecta, $c$ is the speed of light and $v_{\exp}$ is the mean expansion velocity. To further simplify the situation, we assume that $\kappa$ remains roughly constant in the rise phase of a SN. If the mean expansion velocity $v_{\exp}$ also does not change significantly, then the UV pulse of iPTF14atg within four days of its explosion and its main radiation peak observed about 20 days after explosion indicate two distinct radioactive layers. The shallow layer above which the ejecta mass is about 4% of the total ejected mass powered the UV pulse. Furthermore, if the radioactive element in the shallowest layer is $^{56}$Ni, then the UV luminosity of $3\times10^{41}\,\mathrm{ergs\ s^{-1}}$ requires a $^{56}$Ni mass of 0.01 solar mass. Such a configuration has been widely discussed in double-detonation models where a carbon-oxygen white dwarf accretes mass from a helium star. After the helium shell on the surface of the white dwarf reaches a critical mass, helium burns in a detonation wave with such force that a detonation is also ignited in the interior of the white dwarf, making a supernova explosion of sub-Chandrasekhar mass. However, nuclear burning on the surface not only makes $^{56}$Ni, but a layer containing iron-group elements. These elements have a vast number of optically thick lines in the UV to effectively reprocess UV photons into longer wavelengths. Therefore we would not have observed the UV pulse at the early time in this scenario[50]. In addition, some double-detonation models predict weak Si II 6355 Å absorption near the SN peak[51], which contradicts the observed deep Si II absorption in iPTF14atg. Therefore we consider this model as not consistent with observations.

A second model is that the emission arises from circumstellar gas around the progenitor. The circumstellar gas is heated up by either high energy photons from the SN shock breakout (Case 1) or the SN shock itself (Case 2). We assume that the circumstellar gas is optically thin and thus the plausible radiation mechanism is bremsstrahlung. Consider a simple model of a sphere of radius $R_s$ that contains pure material with an atomic number $Z$ and a mass number $A$. The material is completely ionized, so the electron density $n_e$ and the ion density $n_i$ are related by $n_e = Zn_i$. The bremsstrahlung luminosity is

$$L_{ff} = 1.4 \times 10^{-27} T^{1/2} n_e n_i Z^2 \times \frac{4\pi}{3} R_s^3,$$

where all physical quantities are in CGS units. We further assume the critical case that the optical depth of the sphere is unity, i.e.,

$$\tau_s = n_e \sigma_T R_s = 1.$$

Then we can derive analytical expressions for the luminosity, the total mass of the sphere $M$ and the thermal energy $Q$ in terms of $A$, $Z$, $R_s$, and temperature $T$, i.e.,

$$L_{ff} = 5.0 \times 10^{41} Z R_{17} T_5^{1/2} \text{ ergs s}^{-1},$$

$$M = \frac{4\pi}{3} R_s^3 \times n_i A u = 53 \frac{A}{Z} R_{17}^2 M_{sun},$$

$$Q = \frac{4\pi}{3} R_s^3 \times \left(n_i + n_e\right) \times \frac{3}{2} k_B T = 1.3 \times 10^{48} \times \frac{Z+1}{Z} \times R_{17}^2 T_5 \text{ ergs},$$

where $R_s = R_{17} \times 10^{17}\,\text{cm}$, $T = T_5 \times 10^5\,\text{K}$, and $u$ is the atomic mass unit. Because no hydrogen is seen in the iPTF14atg spectra, we assume that the sphere is dominated by helium, then $A = 4$ and $Z = 2$.

In the first case, the circumstellar gas is heated up by the high energy photons from the SN shock breakout. The temperature of the gas is roughly 11,000 K as determined by the optical-UV SED. In order to account for a UV luminosity of $3 \times 10^{41}$ erg s$^{-1}$, the radius $R_s$ has to be as large as $3 \times 10^{16}$ cm. The total mass of the sphere would be ten solar mass and the total thermal energy $10^{47}$ ergs. If the optical depth of the sphere is larger than unity, then we will end up with an even more massive sphere. So we are forced to invoke a sphere containing a mass much larger than a typical SN Ia. Absence of strong Na ID lines also argue against such a massive circumstellar material. In addition, the elliptical host galaxy with no star forming activity also excludes existence of massive stars.

In the second case, the circumstellar gas is ionized by the SN shock. The SN shock has a typical velocity between 20,000 km $s^{-1}$ and 5,000 km $s^{-1}$. Hence, in four days of the SN explosion, the SN shock traveled to $R_s \sim 10^{15} cm$. To account for the UV pulse, this small radius then requires an extremely high temperature of $10^7$ K, which is inconsistent with the observed SED. Therefore we discard this model.

**4 SN specification**

In this section, we perform comparative analysis among iPTF14atg, SN1991bg, SN2002cx and SN2002es on the photometric and spectroscopic evolution and host galaxy and demonstrate that iPTF14atg is likely to belong to the SN2002es family.

**Photometry** The multiband lightcurve of iPTF14atg is shown in Figure 3. Note that there is a $\approx 0.2$ mag difference between the PTF *r*-band and LCOGT *r*-band magnitudes. We calculated synthetic photometry using the iPTF14atg spectra and the filter transmission curves and found that this difference was mainly due to the filter difference.

Because iPTF14atg is not a normal SN Ia, the usual lightcurve fitting tools for normal SNe Ia (e.g., SALT2[52], SNooPy2[53]) are not suitable to determine the lightcurve features. Thus we fit a 5th-order polynomial to the *B*-band lightcurve and derived a B-band peak magnitude of 17.1 mag on May 22.15 and $\Delta m_{15}$ = 1.2 mag. We also infer that the line-of-sight extinction is low because the Galactic extinction in this direction is $A_B$ = 0.032 and because we do not see any sign of strong Na ID absorption in all of our low-resolution and medium-resolution spectra of iPTF14atg. Hence, given the host galaxy distance modulus of 34.9 mag, we conclude that iPTF14atg has an absolute peak magnitude of -17.8 mag and that iPTF14atg is a subluminous outlier of the well-established relation between the peak magnitude and $\Delta m_{15}$[54].

We compare iPTF14atg with the three major families of subluminous SNe Ia with prototypical events of SN1991bg, SN2002cx and SN2002es. From Extended Data Figure 1, we can see that: (1) the peak magnitude of iPTF14atg is brighter than that of SN1991bg, similar to SN2005hk (a typical SN2002cx-family event), and fainter than SN2002es. However, both SN2002cx and SN2002es families have large ranges of peak magnitudes[18, 20]. (2) iPTF14atg evolves more slowly than SN1991bg in both rise and decline phases. (3) iPTF14atg has a slower rise than SN2005hk. (4) Unlike SN2002es, iPTF14atg does not have a break in the lightcurve about 30 days after the peak. A caveat about this comparison is that the lightcurve of iPTF14atg, especially the very early part, might be distorted by the SN-companion collision.

We present the near-UV and optical color evolution of iPTF14atg in Extended Data Figure 2 and compare it with SN2011fe (also known as PTF11kly, a normal SN Ia in M101 six Mpc away from the earth)[13, 55], SN2002es, SN2005hk and SN1991bg. The figure shows that iPTF14atg was initially bluer in *uvm*2-*uvw*1 by more than two magnitudes than SN2011fe which is classified as a near-UV blue event[13]. Though SN2011fe gradually becomes bluer while approaching its peak, iPTF14atg remains the same color and still bluer at peak than SN2011fe by one magnitude. The optical color, indicated by *B-V*, iPTF14atg was initially red, then it quickly became blue in a few days and then followed evolution of SN2002es. Though SN2011fe was also red initially, it gradually became blue during the SN rise, reached its bluest color near the SN peak and turned red later on.

**Spectroscopy** Spectral evolution of iPTF14atg is presented in Extended Data Figure 3 and is compared with those of SN1991bg, SN2005hk and SN2002es in Extended Data Figure 4.

On May 6 (within four days after explosion) when the UV excess was detected, the spectrum of iPTF14atg consisted of a blue continuum superposed by some weak and broad absorption features. In Figure 2 and Extended Data Figure 3 we tentatively identified Si II, S II and Ca II. Combining the photometry from *Swift*/UVOT, the SED can be approximated by a blackbody spectrum of temperature 11, 000 K and radius 6, 000$R$ (Figure 2). None of the known subluminous Type Ia supernovae have been observed at such early time to compare with, so we turned to SN2011fe[2]. Unlike iPTF14atg, the spectra of SN2011fe taken within two days after explosion show clear absorption features commonly seen in a pre-max SN Ia, such as Si II, S II, Mg II, O I and Ca II. We therefore suggest that this spectrum of iPTF14atg has a dominant thermal component from the SN-companion interaction and a weak SN component from intact regions of the SN photosphere. In the next spectrum taken three days later, spectral features like Si II and Ca II have emerged. In the spectrum taken on May 11, we clearly identified Si II around 6100 Å. with its minimum at a velocity of 10,000 km $s^{-1}$. This velocity is lower than that of a normal SN Ia at a similar phase[56].

In the spectrum taken on May 15 (about a week before its maximum brightness; first panel of Extended Data Figure 4), we identified absorption features like Si II, the S II "W" around 5000 Å and O I and concluded that iPTF14atg is a Type Ia SN based on the presence of Si II and S II. The whole spectrum matched well to the SN1991bg spectrum at a similar phase after we redshifted the SN1991bg spectrum by 3,000 km $s^{-1}$. The difference was that iPTF14atg did not have a Ti trough around 4200 Å as deep as SN1991bg. This iPTF14atg spectrum showed little similarity to that of SN2005hk, a typical SN2002cx-like event.

Near the SN peak (second panel of Extended Data Figure 4), the spectrum of iPTF14atg shared similar absorption features with all three families of subluminous Type Ia SNe, though the depth of the absorptions and the continuum shapes differed among them. iPTF14atg had a velocity lower than SN1991bg but higher than SN2005hk. The best match to the overall spectral shape was between iPTF14atg and SN2002es.

The post-maximum spectral evolution (third to fifth panels of Extended Data Figure 4) shows that iPTF14atg shares many spectral similarities with SN2002cx-like events, but differences in the near infrared part of the late-time spectrum taken two months after the SN peak (fifth panel of Extended Data Figure 4) disfavored its classification of a SN2002cx-family. In contrast, iPTF14atg spectra matched well with the limited spectral information available for SN2002es-like events.

Another interesting feature of iPTF14atg is strong and persistent C II 6580 Å absorption. The absorption feature C II, is sometimes seen in pre-max spectra of normal SNe Ia[57, 58] and always disappears before the SN peak. Only in very few cases[56, 59] did the carbon absorption feature exist at a velocity decreasing from 11,000 km $s^{-1}$ to 9,000 km $s^{-1}$ after maximum light. It also has been reported in pre-max or max spectra of SN2002cx-like events but not in post-max spectra[60, 61]. SN2002es may show very weak C II three days after maximum[19]. In the case of iPTF14atg, the C II 6580 Å absorption is first seen at a velocity of 11,000 km $s^{-1}$ in the spectrum taken on May 11 (about 12 days before maximum light). Its velocity decreased to 6, 000 km $s^{-1}$ near the maximum light. After the maximum light, its velocity kept decreasing. It was detected at a velocity of 4,000 km $s^{-1}$ in the spectrum taken on June 6 (two weeks after maximum light) but not later. The long-lasting carbon feature indicates that both high-velocity shallow layers and low-velocity deep layers in the iPTF14atg explosion are carbon-rich, which is evidence for incomplete burning extending deep into the ejecta. The incomplete burning is consistent with a pure deflagration[62].

**Host Galaxy** The host galaxy IC831 is morphologically classified as an elliptical galaxy[63] or an S0 galaxy[64]. Its SDSS spectrum shows very weak H$\alpha$ and [O III] emission. This suggests that the host galaxy has little star forming activity. iPTF14atg occurred about 4.6 kiloparsecs from the center of IC831. In all iPTF14atg spectra, we did not detect any Halpha emission either at the SN location or underlying in the galaxy background, suggesting that iPTF14atg was born in an old population. This strongly argues against that iPTF14atg is a core-collapse SN. Furthermore, SN2002cx-like events prefer star-forming regions while SN2002es-like events are all found in the passive galaxies[18, 20]. Hence the nature of its host galaxy makes iPTF14atg more likely to belong to the SN2002es family.

30. Oke, J. B. & Gunn, J. E. An Efficient Low Resolution and Moderate Resolution Spectrograph for the Hale Telescope. *Publ. Astron. Soc. Pac.* **94**, 586 (1982).

31. Faber, S. M. *et al.* The DEIMOS spectrograph for the Keck II Telescope: integration and testing. In Iye, M. & Moorwood, A. F. M. (eds.) *Instrument Design and Performance for Optical/Infrared Ground-based Telescopes*, vol. 4841 of *Society of Photo-Optical Instrumentation Engineers (SPIE) Conference Series*, 1657–1669 (2003).

32. Oke, J. B. *et al.* The Keck Low-Resolution Imaging Spectrometer. *Publ. Astron. Soc. Pac.* **107**, 375 (1995).

33. Hook, I. M. *et al.* The Gemini-North Multi-Object Spectrograph: Performance in Imaging, Long-Slit, and Multi-Object Spectroscopic Modes. *Publ. Astron. Soc. Pac.* **116**, 425–440 (2004).

34. Brink, H. *et al.* Using machine learning for discovery in synoptic survey imaging data. *Mon. Not. Astron. Roy. Soc.* **435**, 1047–1060 (2013).

35. Wozniak, P. R. *et al.* Automated Variability Selection in Time-domain Imaging Surveys Using Sparse Representations with Learned Dictionaries. In *American Astronomical Society Meeting Abstracts #221*, vol. 221 of *American Astronomical Society Meeting Abstracts*, 431.05 (2013).

36. Bue, B. D., Wagstaff, K. L., Rebbapragada, U. D., Thompson, D. R. & Tang, B. Astronomical data triage for rapid science return. In *Proceedings of the 2014 conference on Big Data from Space (BiDS'14)* (2014).

37. Gal-Yam, A. *et al.* Real-time Detection and Rapid Multiwavelength Follow-up Observations of a Highly Subluminous Type II-P Supernova from the Palomar Transient Factory Survey. *Astrophys. J.* **736**, 159 (2011).

38. Holoien, T. W.-S. *et al.* ASAS-SN Discoveries of a Probable Supernova in IC 0831 and a Possible Extreme (delta V > 6.6 mag) M-dwarf Flare. *The Astronomer's Telegram* **6168**, 1 (2014).

39. Wagner, R. M. *et al.* Spectroscopic Classification of ASASSN-14bd. *The Astronomer's Telegram* **6203**, 1 (2014).

40. Bolton, A. S. *et al.* Spectral Classification and Redshift Measurement for the SDSS-III Baryon Oscillation Spectroscopic Survey. *Astron. J.* **144**, 144 (2012).

41. Planck Collaboration *et al.* Planck 2013 results. XVI. Cosmological parameters. *Astron. & Astrophys.* **571**, 16 (2014).

42. Yaron, O. & Gal-Yam, A. WISeREP - An Interactive Supernova Data Repository. *Publ. Astron. Soc. Pac.* **124**, 668–681 (2012).

43. Roming, P. W. A. *et al.* The Swift Ultra-Violet/Optical Telescope. In Flanagan, K. A. & Siegmund, O. H. W. (eds.) *X-Ray and Gamma-Ray Instrumentation for Astronomy XIII*, vol. 5165 of *Society of Photo-Optical Instrumentation Engineers (SPIE) Conference Series*, 262– 276 (2004).

44. Burrows, D. N. *et al.* The Swift X-Ray Telescope. In Flanagan, K. A. & Siegmund, O. H. W. (eds.) *X-Ray and Gamma-Ray Instrumentation for Astronomy XIII*, vol. 5165 of *Society of Photo-Optical Instrumentation Engineers (SPIE) Conference Series*, 201–216 (2004).

45. The HEASoft software can be downloaded at http://heasarc.nasa.gov/heasoft/.

46. Breeveld, A. A. *et al.* An Updated Ultraviolet Calibration for the Swift/UVOT. In McEnery, J. E., Racusin, J. L. & Gehrels, N. (eds.) *American Institute of Physics*

*Conference Series*, vol. 1358 of *American Institute of Physics Conference Series*, 373–376 (2011). 1102.4717.

47. WebPIMMS is available at http://heasarc.gsfc.nasa.gov/cgi-bin/Tools/w3pimms/w3pimms.pl.

48. Kalberla, P. M. W. *et al.* The Leiden/Argentine/Bonn (LAB) Survey of Galactic HI. Final data release of the combined LDS and IAR surveys with improved stray-radiation corrections. *Astron. & Astrophys.* **440**, 775–782 (2005).

49. Ofek, E. O. *et al.* The Palomar Transient Factory Photometric Calibration. *Publications of the Astronomical Society of the Pacific* **124**, 62–73 (2012).

50. Kromer, M. *et al.* Double-detonation Sub-Chandrasekhar Supernovae: Synthetic Observables for Minimum Helium Shell Mass Models. *Astrophys. J.* **719**, 1067–1082 (2010).

51. Nugent, P., Baron, E., Branch, D., Fisher, A. & Hauschildt, P. H. Synthetic Spectra of Hydrodynamic Models of Type Ia Supernovae. *Astrophys. J.* **485**, 812–819 (1997).

52. Guy, J. *et al.* SALT2: using distant supernovae to improve the use of type Ia supernovae as distance indicators. *Astron. & Astrophys.* **466**, 11–21 (2007).

53. Burns, C. R. *et al.* The Carnegie Supernova Project: Light-curve Fitting with SNooPy. *Astron. J.* **141**, 19 (2011).

54. Phillips, M. M. The absolute magnitudes of Type IA supernovae. *Astrophys. J. Lett.* **413**, L105–L108 (1993).

55. Vinkó, J. *et al.* Testing supernovae Ia distance measurement methods with SN 2011fe. *Astron. & Astrophys.* **546**, A12 (2012).

56. Parrent, J. T. *et al.* Analysis of the Early-time Optical Spectra of SN 2011fe in M101. *Astrophys. J.* **752**, L26 (2012).

57. Thomas, R. C. *et al.* Type Ia Supernova Carbon Footprints. *Astrophys. J.* **743**, 27 (2011).

58. Silverman, J. M. & Filippenko, A. V. Berkeley Supernova Ia Program - IV. Carbon detection in early-time optical spectra of Type Ia supernovae. *Mon. Not. Astron. Roy. Soc.* **425**, 1917–1933 (2012).

59. Cartier, R. *et al.* Persistent C II Absorption in the Normal Type Ia Supernova 2002fk. *Astrophys. J.* **789**, 89 (2014).

60. Foley, R. J. *et al.* Early- and Late-Time Observations of SN 2008ha: Additional Constraints for the Progenitor and Explosion. *Astrophys. J.* **708**, L61–L65 (2010).

61. Parrent, J. T. *et al.* A Study of Carbon Features in Type Ia Supernova Spectra. *Astrophys. J.* **732**, 30 (2011).

62. Gamezo, V. N., Khokhlov, A. M., Oran, E. S., Chtchelkanova, A. Y. & Rosenberg, R. O. Thermonuclear Supernovae: Simulations of the Deflagration Stage and Their Implications. *Science* **299**, 77–81 (2003).

63. Scodeggio, M., Giovanelli, R. & Haynes, M. P. The Universality of the Fundamental Plane of E and S0 Galaxies: Sample Definition and I-Band Photometric Data. *Astron. J.* **116**, 2728– 2737 (1998).

64. Huchra, J. P. *et al.* The 2MASS Redshift Survey-Description and Data Release. *Astrophys. J. Suppl.* **199**, 26 (2012).

65. The Nugent SN template is available at https://c3.lbl.gov/nugent/nugent_templates.html.

**Table Extended Data Table 1: Spectroscopic observation log**

| Date (UT) | Telescope/Instrument | $\Delta\lambda$ (Å) | Wavelength (Å) | Observer | Data Reducer |
|---|---|---|---|---|---|
| May 6.32 | ARC-3.5m/DIS[a] | 10 | 3500 - 9500 | Cao | Cao |
| May 6.96 | NOT/ALFOSC[b] | 16.2 | 3500 - 9000 | O. Smirnova | Taddia |
| May 9.25 | ARC-3.5m/DIS[a] | 10 | 3500 - 9500 | Kasliwal | Cao |
| May 11.04 | NOT/ALFOSC[b] | 16.2 | 3500 - 9000 | Y. F. Martinez | Taddia |
| May 15.96 | NOT/ALFOSC[b] | 16.2 | 3500 - 9000 | A. Nyholm | S. Papadogiannakis |
| May 21.31 | ARC-3.5m/DIS[a] | 10 | 3500 - 9500 | Cao | Cao |
| May 24.21 | Hale/DBSP[c] | 10 | 3300 - 10000 | A. Waszczak | A. Rubin & O. Yaron |
| May 26.35 | Keck-II/DEIMOS[d] | 1.5 | 5700 - 8200 | Cao | A. De Cia |
| May 28.33 | Keck-I/LRIS[e] | 7 | 3300 - 10000 | D. A. Perley | D. A. Perley |
| June 3.15 | ARC-3.5m/DIS[a] | 10 | 3500 - 9500 | Cao | Cao |
| June 6.23 | Hale/DBSP[c] | 10 | 3300 - 10000 | A. Waszczak | O.Yaron |
| June 29.30 | Keck-I/LRIS[e] | 7 | 3300 - 5500 | Cao & G. E. Duggan | D. A. Perley |
| | | 4.7 | 5800 - 7400 | | |
| July 30.24 | Keck-I/LRIS[e] | 4 | 3300 - 5500 | Cao | D. A. Perley |
| | | 2.5 | 5400 - 7000 | | |
| August 20.24 | Gemini-N/GMOS[f] | 3 | 4000 - 9000 | | Kasliwal |

[a] The Dual Image Spectrograph (DIS) on the ARC-3.5m telescope at the Apache Observatory, New Mexico, USA.

[b] The Andalucia Faint Object Spectrograph and Camera (ALFOSC) on the Nordic Optical Telescope at La Palma, Spain.

[c] The Double Spectrograph (DBSP)[30] on the Palomar 200-inch Hale telescope at Palomar Observatory, California, USA.

[d] The DEep Imaging Multi-Object Spectrograph (DEIMOS)[31] on the Keck-II telescope at Mauna Kea, Hawaii, USA.

[e] The Low Resolution Imaging Spectrometer (LRIS)[32] on the Keck-I telescope at Mauna Kea, Hawaii, USA.

[f] The Gemini Multi-Object Spectrograph (GMOS)[33] on the Gemini-N telescope at Mauna Kea, Hawaii, USA.

## Table Extended Data Table 2: Swift Observation of iPTF14atg

| UT Time | UVOT (counts/sec)[a] | | | | | | XRT (counts/sec)[b] |
|---|---|---|---|---|---|---|---|
| | uvw2 | uvm2 | uvw1 | u | b | v | |
| May 06.67 - 06.74 | $0.297 \pm 0.028$ | $0.176 \pm 0.014$ | $0.399 \pm 0.044$ | $1.251 \pm 0.116$ | $2.354 \pm 0.168$ | $1.491 \pm 0.131$ | $< 3.7 \times 10^{-3}$ |
| May 08.53 - 08.61 | $0.112 \pm 0.019$ | $0.096 \pm 0.014$ | $0.324 \pm 0.041$ | $1.374 \pm 0.151$ | $2.444 \pm 0.212$ | $1.472 \pm 0.163$ | $< 4.9 \times 10^{-3}$ |
| May 12.98 - 13.12 | $0.208 \pm 0.019$ | $0.182 \pm 0.019$ | $0.578 \pm 0.044$ | $2.971 \pm 0.152$ | $4.515 \pm 0.194$ | $2.288 \pm 0.136$ | $< 5.2 \times 10^{-3}$ |
| May 15.25 - 15.38 | $0.351 \pm 0.057$ | $0.296 \pm 0.041$ | $0.681 \pm 0.100$ | $3.636 \pm 0.414$ | $5.557 \pm 0.531$ | $2.945 \pm 0.371$ | $< 2.2 \times 10^{-3}$ |
| May 18.99 - 19.05 | $0.452 \pm 0.051$ | $0.297 \pm 0.024$ | $1.081 \pm 0.094$ | $5.381 \pm 0.372$ | $6.600 \pm 0.417$ | $3.580 \pm 0.285$ | $< 8.8 \times 10^{-3}$ |
| May 20.31 - 20.46 | $0.404 \pm 0.034$ | $0.325 \pm 0.024$ | $0.956 \pm 0.067$ | $5.758 \pm 0.307$ | $7.084 \pm 0.347$ | $3.178 \pm 0.214$ | $< 6.5 \times 10^{-3}$ |
| May 21.45 - 21.46 | $0.437 \pm 0.067$ | $0.262 \pm 0.042$ | $0.809 \pm 0.115$ | $6.015 \pm 0.578$ | $7.454 \pm 0.669$ | $3.178 \pm 0.407$ | $< 2.4 \times 10^{-2}$ |
| May 25.45 - 25.65 | $0.166 \pm 0.019$ | $0.156 \pm 0.014$ | $0.483 \pm 0.041$ | $4.228 \pm 0.219$ | $6.808 \pm 0.291$ | $3.590 \pm 0.195$ | $< 3.6 \times 10^{-3}$ |
| May 27.65 - 27.85 | $0.124 \pm 0.023$ | $0.084 \pm 0.014$ | $0.440 \pm 0.053$ | $3.595 \pm 0.274$ | $6.804 \pm 0.400$ | $3.494 \pm 0.266$ | $< 6.8 \times 10^{-3}$ |
| May 30.24 - 30.52 | $0.078 \pm 0.018$ | $0.029 \pm 0.009$ | $0.352 \pm 0.045$ | $2.551 \pm 0.227$ | $4.883 \pm 0.330$ | $3.490 \pm 0.257$ | $< 7.0 \times 10^{-3}$ |
| Jun 07.38 - 07.65 | $0.041 \pm 0.011$ | $0.019 \pm 0.006$ | $0.187 \pm 0.028$ | $1.101 \pm 0.119$ | $2.936 \pm 0.195$ | $2.279 \pm 0.162$ | $< 3.8 \times 10^{-3}$ |
| Jun 17.39 - 17.60 | $0.052 \pm 0.014$ | $0.027 \pm 0.009$ | $0.143 \pm 0.023$ | $0.859 \pm 0.102$ | $2.235 \pm 0.168$ | $1.782 \pm 0.196$ | $< 5.4 \times 10^{-3}$ |
| Jun 21.37 - 21.45 | $0.039 \pm 0.011$ | $0.023 \pm 0.006$ | $0.127 \pm 0.024$ | $0.833 \pm 0.107$ | $2.377 \pm 0.176$ | $1.851 \pm 0.143$ | $< 4.9 \times 10^{-3}$ |
| Jun 25.24 - 25.53 | $0.045 \pm 0.012$ | $0.018 \pm 0.007$ | $0.120 \pm 0.026$ | $0.801 \pm 0.113$ | $1.624 \pm 0.153$ | $0.413 \pm 0.039$ | $< 4.6 \times 10^{-3}$ |
| Nov 12.04 - 12.11[c] | $0.004 \pm 0.010$ | $0.017 \pm 0.008$ | $0.080 \pm 0.025$ | $0.559 \pm 0.115$ | $1.576 \pm 0.189$ | $0.953 \pm 0.156$ | |

[a] The uncertainties are at a 66.8% CL.

[b] The upper limits are at a 99.7% CL.

[c] This is a reference epoch to remove host galaxy contamination.

**Extended Data Figure 1: Comparative analysis of iPTF14atg lightcurve.** The lightcurves of iPTF14atg are compared to SN1991bg[65], a typical SN2002cx-like event SN2005hk, and SN2002es. The red triangles are upper limits at a 99.9999% CL for non-detections of iPTF14atg. The error bars denote 1-σ uncertainties.

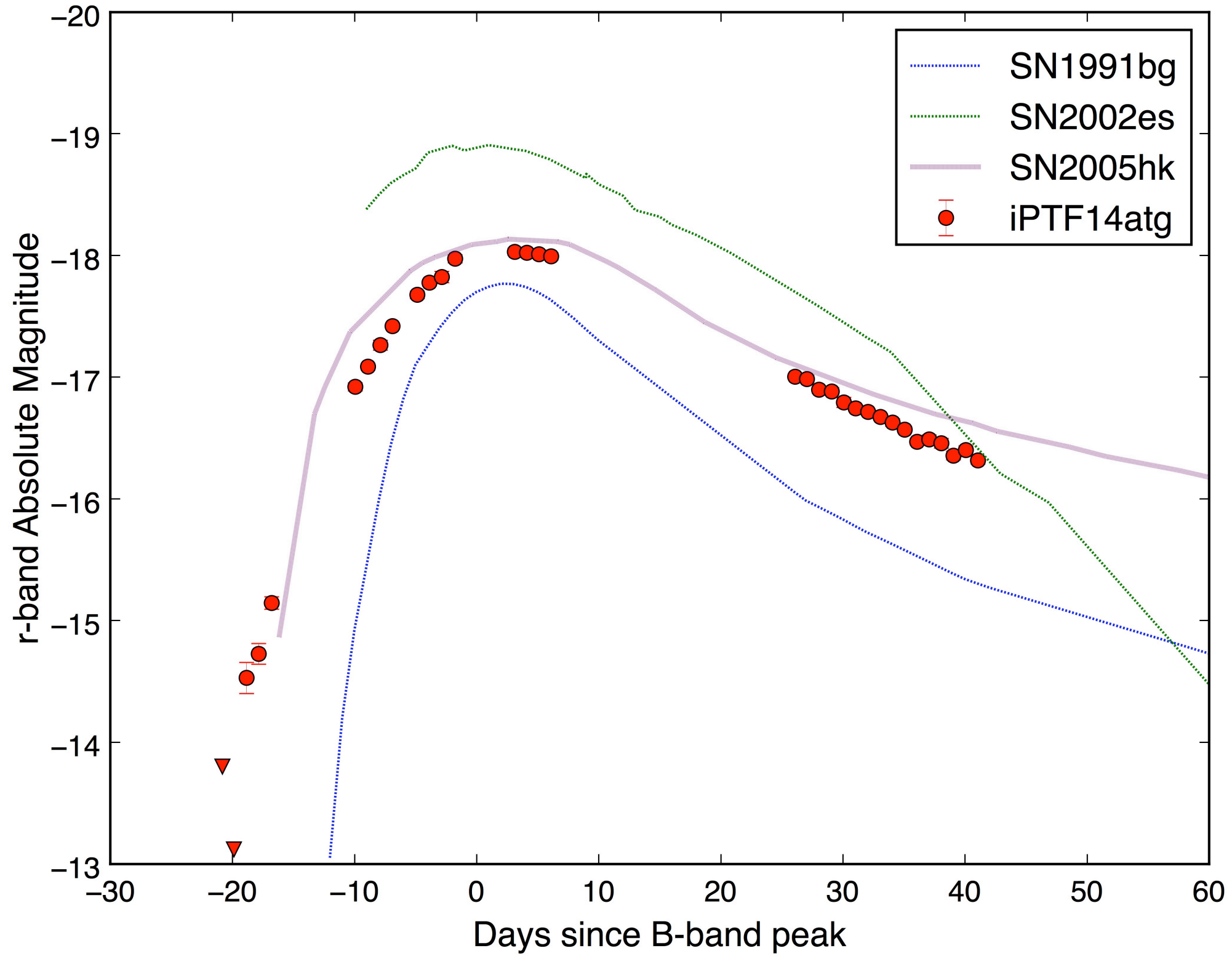

**Extended Data Figure 2: Comparative analysis of iPTF14atg color evolution.** The color curves of iPTF14atg are compared to SN1991bg, SN2005hk, SN2002es and a normal event SN2011fe. The error bars denote 1-σ uncertainties.

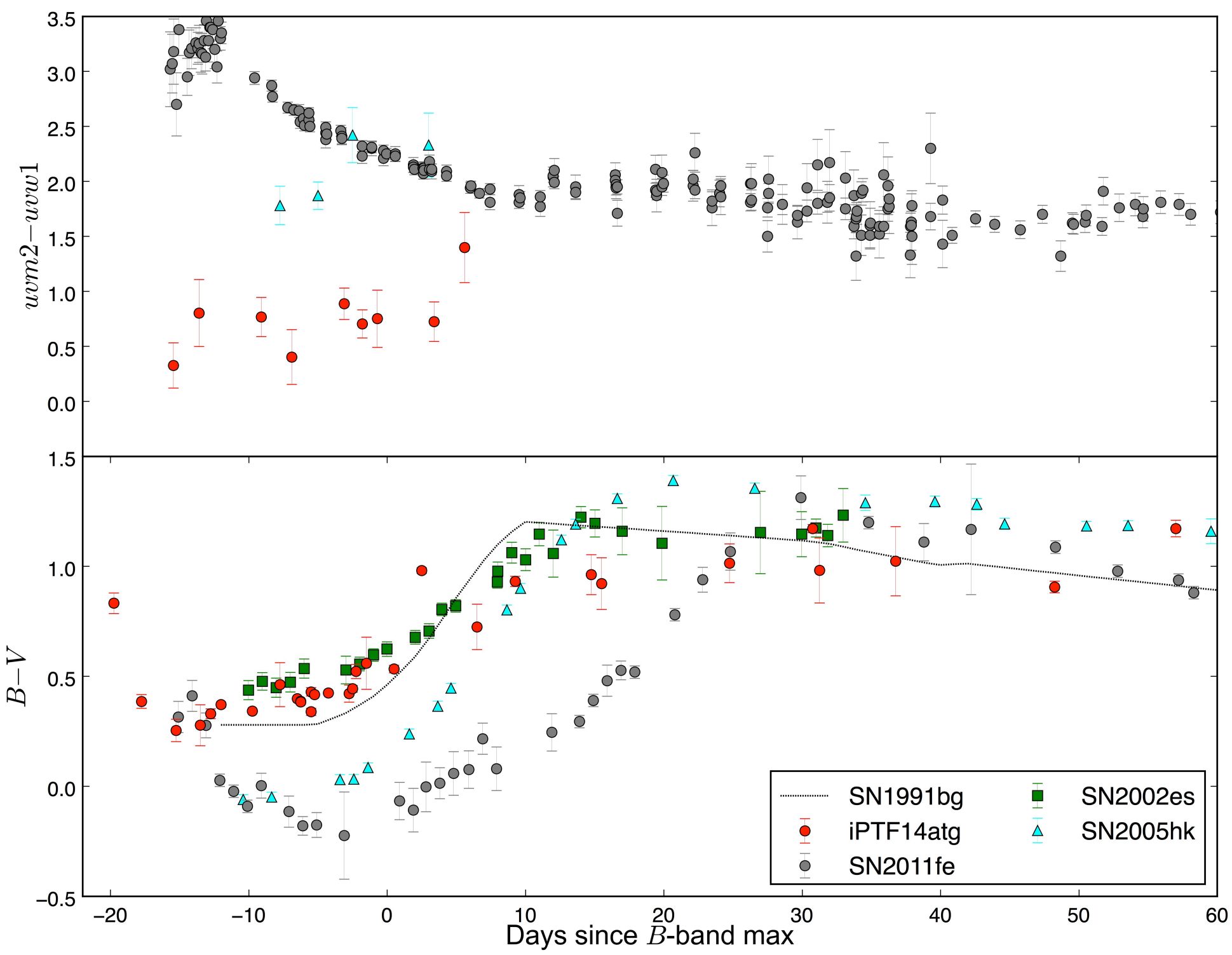

**Extended Data Figure 3: The spectral evolution of iPTF14atg.** Ticks at the top of the figure label major absorption features.

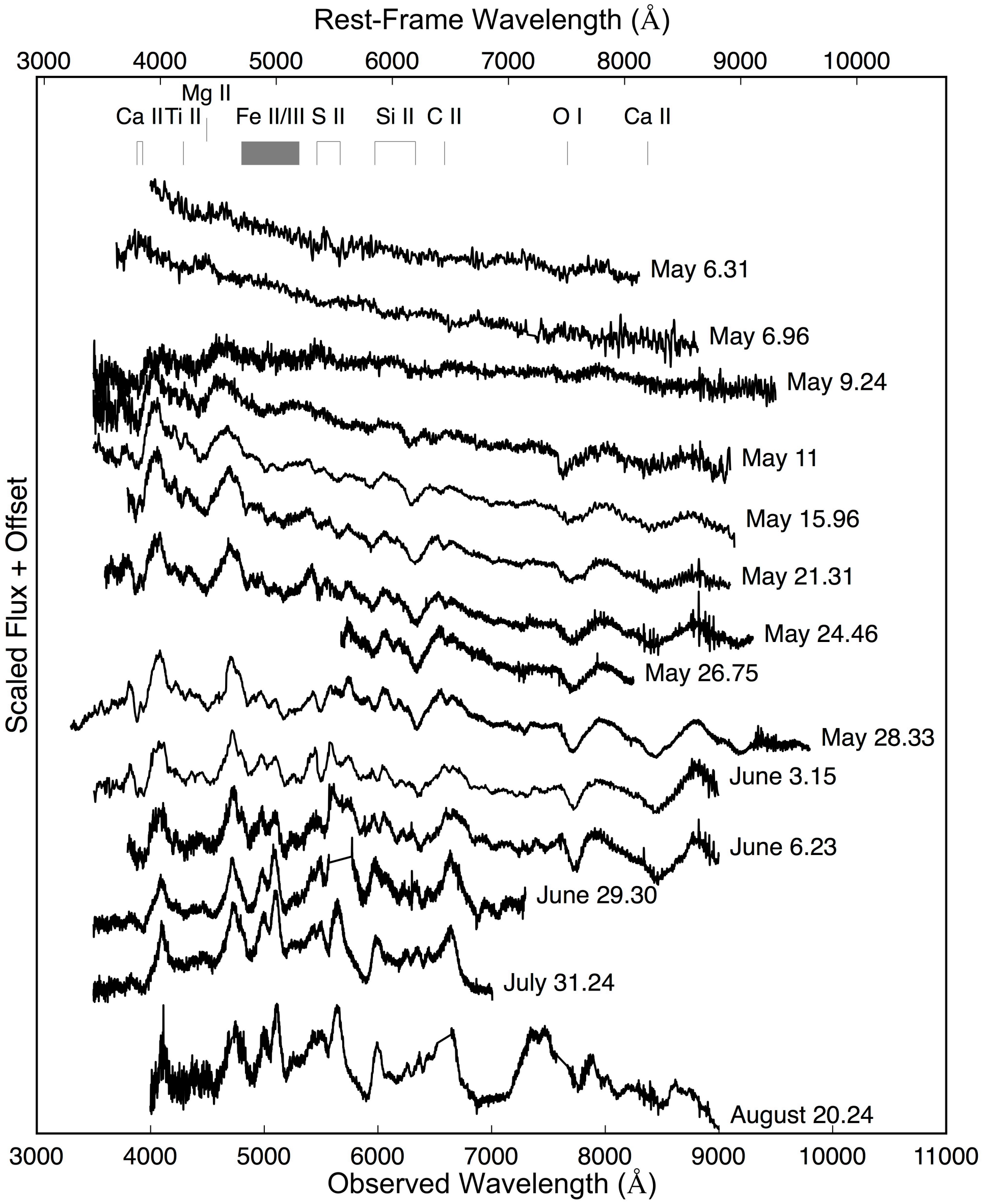

**Extended Data Figure 4: Comparative analysis of iPTF14atg spectra** The spectra of iPTF14atg at different phases are compared with those of SN1991bg, SN2005bl (SN1991bg-like), SN2005hk, SN2002es and PTF10ujn (SN2002es-like) at similar phases.

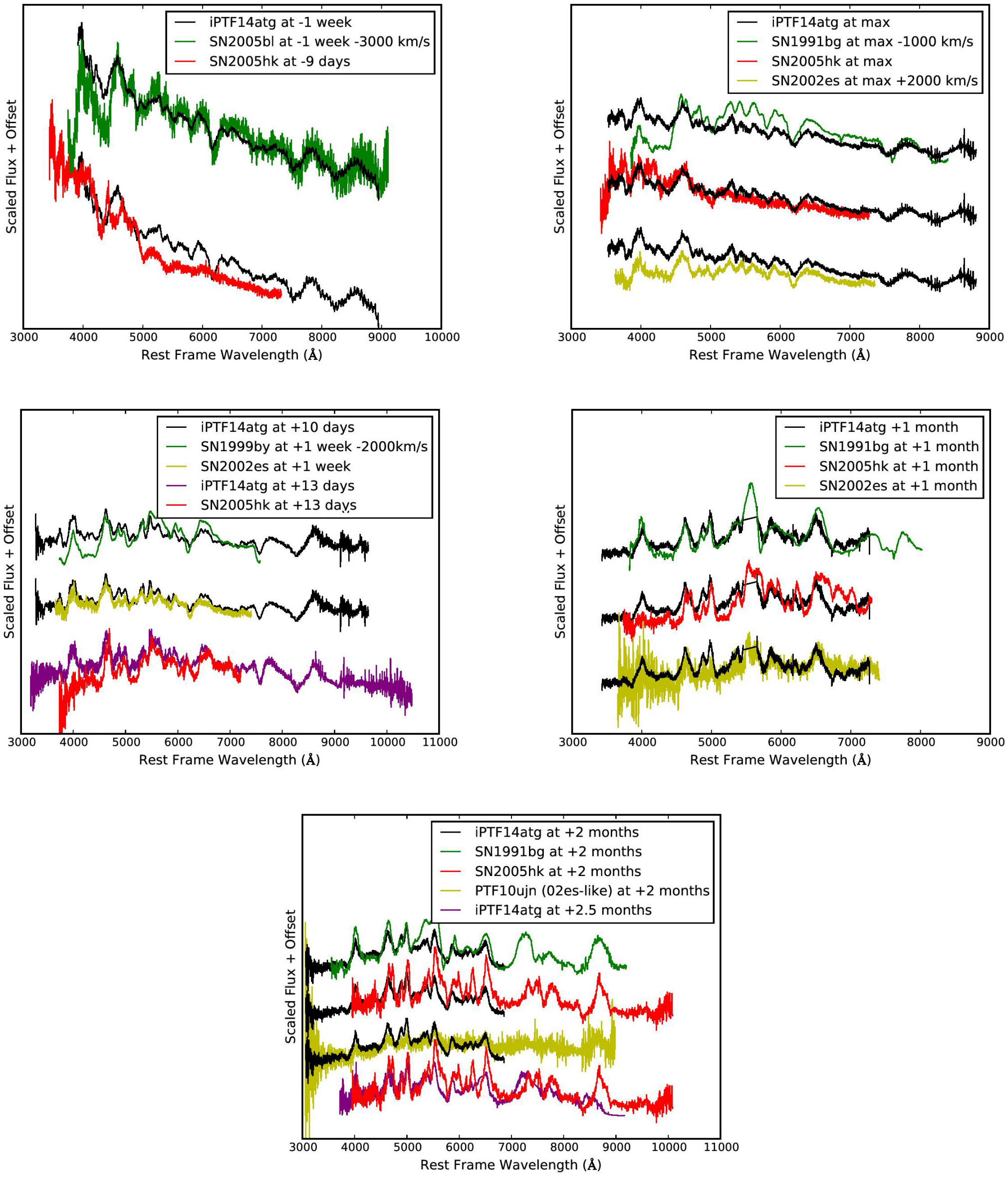